# Evolutionary optimization of all-dielectric magnetic nanoantennas

*Nicolas Bonod∥, Sébastien Bidault°, Geoffrey W. Burr#, and Mathieu Mivelle‡\**

∥Aix Marseille Univ, CNRS, Centrale Marseille, Institut Fresnel, Marseille, France

°ESPCI Paris, PSL Research University, CNRS, Institut Langevin, 75005 Paris, France

#IBM Almaden Research Center, San Jose, California 95120, United States

‡Sorbonne Université, CNRS, Institut des NanoSciences de Paris, UMR 7588, 75005 Paris, France

Abstract:

Magnetic light and matter interactions are generally too weak to be detected, studied and applied technologically. However, if one can increase the magnetic power density of light by several orders of magnitude, the coupling between magnetic light and matter could become of the same order of magnitude as the coupling with its electric counterpart. For that purpose, photonic nanoantennas have been proposed, and in particular dielectric nanostructures, to engineer strong local magnetic field and therefore increase the probability of magnetic interactions. Unfortunately, dielectric designs suffer from physical limitations that confine the magnetic hot spot in the core of the material itself, preventing experimental and technological implementations. Here, we demonstrate that evolutionary algorithms can overcome such limitations by designing new dielectric photonic nanoantennas, able to increase and extract the optical magnetic field from high refractive index materials. We also demonstrate that the magnetic power density in an evolutionary optimized dielectric nanostructure can be increased by a factor 5 compared to state-of-the-art dielectric nanoantennas. In addition, we show that the fine details of the nanostructure are not critical in reaching these aforementioned features, as long as the general shape of the motif is maintained. This advocates for the feasibility of nanofabricating the optimized antennas experimentally and their subsequent application. By designing all–dielectric magnetic antennas that feature local magnetic hot-spots outside of high refractive index materials, this work highlights the potential of evolutionary methods to fill the gap between electric and magnetic light-matter interactions, opening up new possibilities in many research fields.

Graphic

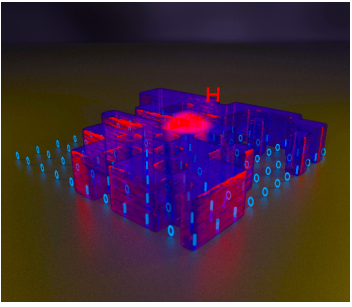

Light and matter interactions are often considered to be mediated by the optical electric field only, neglecting its magnetic counterpart.[1] This is particularly true in quantum optics, where the electric field couples much more strongly, by at least two orders of magnitude, to the electric dipole of a quantum emitter than the magnetic field with the magnetic dipole.[2-5] Nevertheless, if the magnetic energy density of light was increased by several orders of magnitude with respect to the electric one, we would be able to counterbalance the predominance of electric light-matter interactions. Magnetic light and matter interactions could then lead to complete new applications in a variety of fields such as optoelectronics, nonlinear and nano-optics, spintronics, metamaterials, chiral optics,[6] sensing,[7] and photochemistry[8] among others. After experimental demonstrations of the coupling between magnetic dipolar emitters and non-resonant photonic structures,[9-15] both dielectric[5, 16-30] and plasmonic[29, 31-34] optical nanoantennas were recently proposed to boost the magnetic field of light in several theoretical studies. Along the same lines, some of these structures were theoretically shown to strongly increase the emission rates of magnetic dipoles.[5, 19, 21, 24-25, 27-28, 32, 34] In fact, it has been demonstrated experimentally only very recently that optical nanoantennas could manipulate the emission of such dipoles at visible [35-39] or near-infrared[40] wavelengths. In that context, based on Mie resonances, high index dielectric nanoparticles are of particular interest due to their potential to efficiently enhance the magnetic optical field through strong displacement currents taking place inside these nanoantennas.[25] One drawback though, is the fact that the magnetic enhancement occurs inside the material, in the core of the structures, making it difficult to access experimentally.[41] Hollow nano-disks[24, 27-28] (i.e nanocylinders) were proposed to overcome this obstacle, granting access to the magnetic field that would be partially enhanced in air and not entirely in the dielectric material. We recently used this approach to experimentally couple such structures to magnetic emitters, demonstrating the manipulation of the magnetic local density of states.[37] However, it is still very challenging to place a nanoscale piece of material inside hollow nanodisks, making nanostructures featuring easily accessible magnetic hotspots particularly appealing. Along those lines, it was recently demonstrated that evolutionary algorithms[42] can produce optical nanostructures behaving far better than conventional photonic antennas[43-45] or devices.[46-49] By coupling the power of computation algorithms to specific goals and the intuition of researchers, these approaches are redefining what might be the future of nanophotonics.[42]

Objectives:

In this paper, we demonstrate that evolutionary algorithms are a powerful alternative to design dielectric optical nanostructures able to strongly enhance, in the near field, the optical magnetic field of light at visible wavelengths. Furthermore, while dielectric nanoantennas usually increase this magnetic field inside the material (i.e. inside the antenna), the architectures developed here feature a reachable magnetic hot spot just above the antenna, with up to 5 times more enhancement than hollow nanodisks. Finally, we show that, although algorithmically optimized geometries are more elaborate than the shape of conventional nanoantennas, their photonic response depends very weakly on the fine details of the design. In fact, a rough fabrication process following the general shape of the optimized antenna would still lead to more than a 4 times enhancement of the optical magnetic intensity with respect to state-of-the-art nanoantennas, highlighting the feasibility of their experimental implementation.

Results:

Numerical evolutionary algorithm techniques mimic the natural selection processes that happen during the evolution of species. Figure 1a summarizes this approach for the design of silicon-based nanoantennas featuring enhanced magnetic fields: first, a population of random nanostructures is generated; then, each population is evaluated for our specific goal through Finite Difference Time Domain (FDTD) simulations, in order to obtain the highest optical magnetic intensity at a targeted location in the near field of the nanostructure. From this analysis, we select the antennas that provide the best results and we create a new population by either mutations or breeding of these selected designs. This second generation is then evaluated again through FDTD, and so on until, after several generations, an optimized solution emerges. In here, each population is composed of 20 elements and each new generation is elaborated from the 5 best structures favored during the selection process (see supporting information). Each structure element of each generation is reduced to a 11x11 binary matrix made of 0 and 1,[43] as shown in Figure 1b, and each 1 of this matrix corresponds to a parallelepiped made of silicon in the FDTD simulation, the rest being air. Each centre of these silicon blocks is then placed in an array with a periodicity of 30 nm in x and y inside the FDTD simulation. At first, the x and y dimensions of the silicon parallelepipeds are set to 40x40 nm$^2$: this dimension is larger than the periodicity in order to have overlapping blocks of silicon at certain places, allowing displacement currents to take place inside the dielectric material of the antenna. An example of the corresponding structure associated to the matrix in Figure 1b is displayed in Figure 1c. The full width D in Figure 1c was set to 340 nm. The thickness h of the structure was 110 nm (Figure 1d), in order to ensure a π phase shift of the incoming plane wave between the entrance (lower part) and the exit (top part) of the silicon elements (for an optical index of the silicon obtained experimentally[37]). This condition is usually required for dielectric Mie resonators used to enhance the magnetic optical field[41], therefore we took it as a starting point in the simulations. The nanostructure was illuminated in normal incidence by a plane wave, 600 nm in wavelength and circularly polarized. The maximization of the magnetic intensity was then investigated just above (in the first mesh cell), and in the centre of the structure (red spot in Figure 1c and d).

The FDTD method is time consuming and requires high computational power. Therefore, in order to speed up the selection process, we first used a rough discretization mesh of 10x10x10 nm$^3$ size to describe the binary matrices in the simulations. For the same purpose of saving computation time, the simulations were stopped after 10 optical cycles, a number that warrants to reach a plateau in the optical response of the simulated nanostructures.

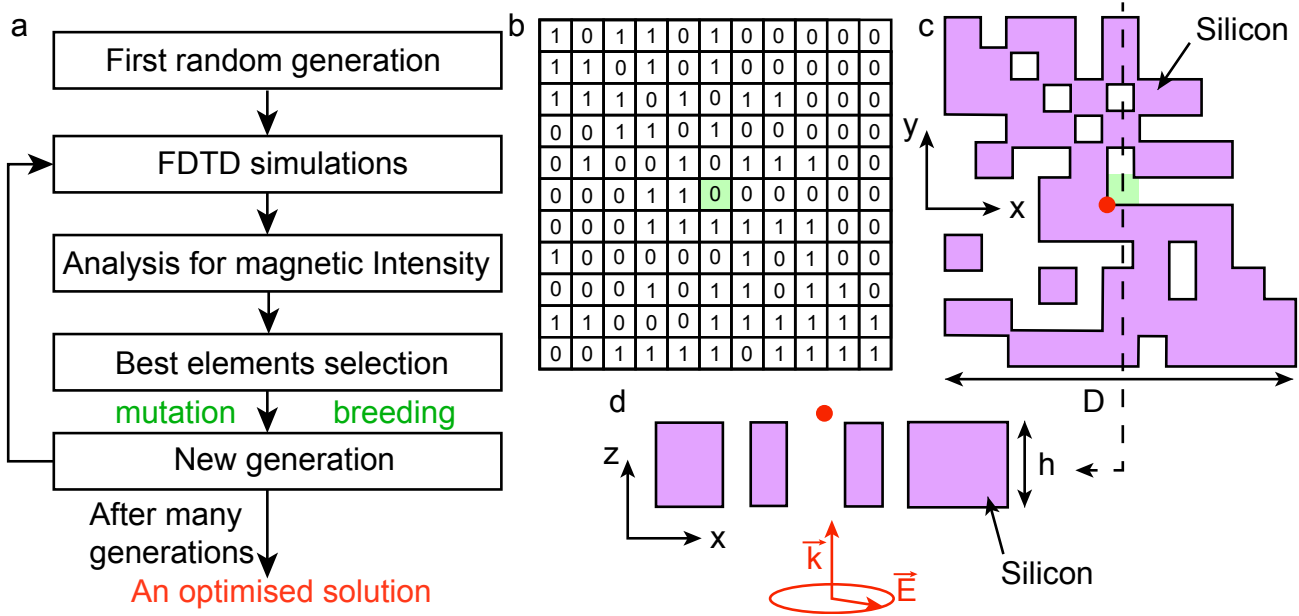

**Figure 1. General concept of the genetic algorithm (GA) developed for designing nanostructures that optimize local magnetic fields. a) Schematic of the different steps realized during the selection process. First a random generation of 20 elements is created, each element is evaluated through Finite Difference Time Domain (FDTD) simulations, the 5 structures giving the highest magnetic field intensity are selected and 20 new elements are created by either mutating or breeding the 5 selected elements. This new generation is then evaluated through FDTD again until after several generations, an optimized solution emerges. b) Typical binary matrix defining a structure to be evaluated: each 1 corresponds to a silicon parallelepiped in the FDTD simulation. c) Schematic representation, in the (x,y) plane, of the structure produced by the matrix in b), (D = 330 nm). d) Cross section of the structure in c), in a (x,z) plane (symbolized by the dashed line in c)), (h = 110 nm). The nanostructures are excited by a plane wave, at λ=600 nm, with a circular polarization and propagating in a direction normal to the (x,y) plane. The red spot in c) and d) corresponds to the position where the magnetic intensity enhancement is evaluated (the first mesh cell above the structure).**

Taking into account all these parameters, figure 2a displays the typical optical magnetic field intensity enhancement in the centre (x=y=D/2, z=h+5 nm) of the generated structures during the evolutionary process. In here, each data point represents a single simulation of a unique nanostructure. Each generation is therefore composed of 20 data points, meaning that more than 7000 independent simulations are presented in figure 2a. As we can see, the magnetic field intensity enhancement quickly increases with the number of generations, reaching a plateau after about 150 successive mutations of the nanostructure geometry. Figures 2b and d show 2 structures at different steps of the evolutionary process, at respectively generation 50 and 100. We observe that the central shape is already defined at a very early stage of the selection process, suggesting an important role in the magnetic field intensity enhancement. In addition, figures 2c and e display the magnetic field intensity enhancement of the structures presented in figures 2b and d, respectively. We notice that the magnetic intensity distributions are rather similar in shape but with a higher confinement and enhancement in the case of generation 100, indicating that the changes at the periphery of the structure from generation 50 to 100 resulted in a high concentration of the magnetic energy density and therefore to a higher intensity enhancement.

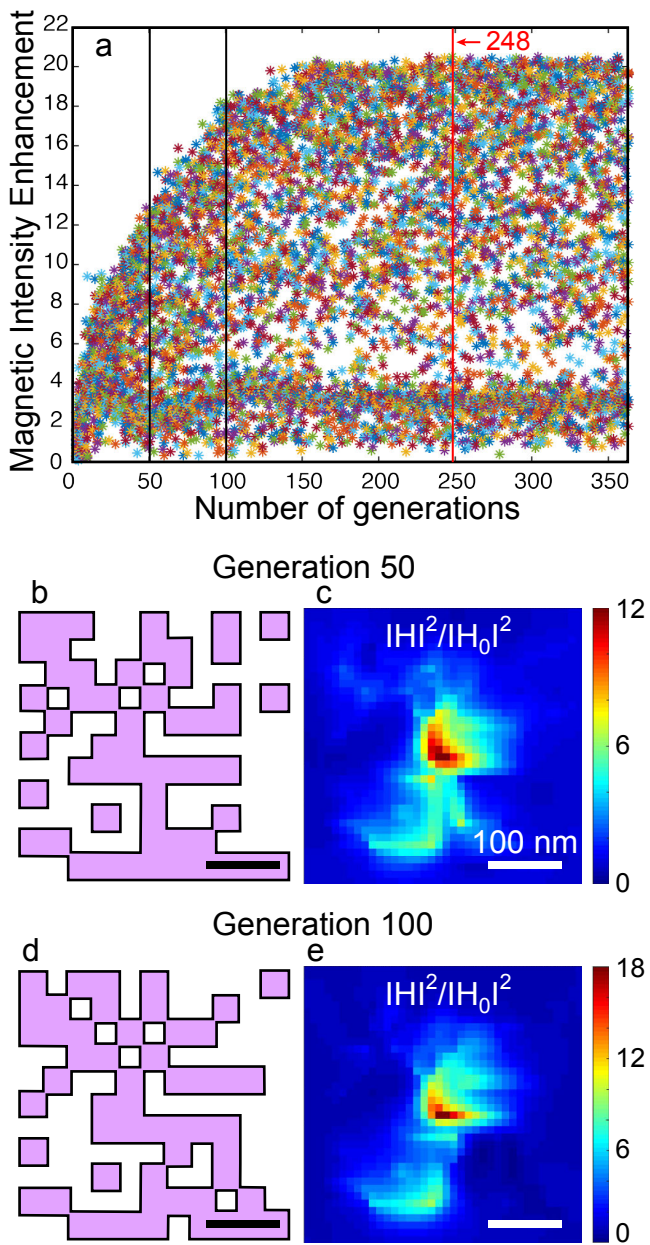

**Figure 2. Magnetic field intensity enhancement during the evolutionary process. a) Magnetic intensity enhancement in the centre of the nanoantenna (red point in figures 1-c-d) versus the number of generations. Each generation displays the results of 20 structures (20 data points), characterizing the spreading of the magnetic intensity increase for each generation. b) Optimal structure at generation 50 and c) corresponding magnetic intensity enhancement distribution in a plane 5 nm above the silicon (z=115nm). d), e) Optimal structure and distribution of the magnetic field intensity increase, in the same plane as c), at generation 100.**

From figure 2a, we were also able to identify the design with the highest magnetic field enhancement selected by the algorithm, which was found to be at generation 248 and is presented in figure 3a. In order to better describe the behavior of that specific structure, the mesh of the simulation was then changed to 2x2x2 nm³ and a second optimization allowed by this fine mesh was performed, consisting in changing the aspect ratio (global size of the antenna), the thickness and the sizes of the parallelepipeds made of silicon (see supporting information). The final structure was then found to be made of parallelepipeds of silicon with dimensions of 36x36 nm$^2$ in x,y and 92 nm in z, surrounded by air and placed with a periodicity of 33 nm in x and y, with a total width D of 366 nm. This nanostructure was then compared, in the same conditions and using the same materials, to a hollow nanodisk,[24, 28] the best silicon nanoantenna known so far with both high enhancement and good accessibility of the field.

Figures 3c and d display the magnetic intensity enhancement 2nm above the GA design and the hollow nanodisk, respectively. Although the reduction of the mesh size slightly lowers the magnetic field intensity enhancement, we can see that the GA reaches a magnetic energy density that is five times higher than that of the hollow nanodisk, demonstrating the potential of evolutionary algorithms to generate accessible hot spots of the magnetic field only. Furthermore, figures 3e and f also display respectively the normalized electric field intensity and the ratio between magnetic and electric normalized intensities in the same plane as figure 3c. From these figures, we see that the electric and magnetic field do not overlap in the near field of our GA design. In addition, we can observe that, at certain positions, the ratio of magnetic over electric intensities is very high (more than 80). Although being not the main focus of this paper, it is worth noticing that such a genetic algorithm could equally be used to optimize this magnetic/electric ratio instead of optimizing the magnetic field enhancement. This would be of particular interest to study specifically the coupling between magnetic light and matter without contamination of the electric optical field and with a very high efficiency.

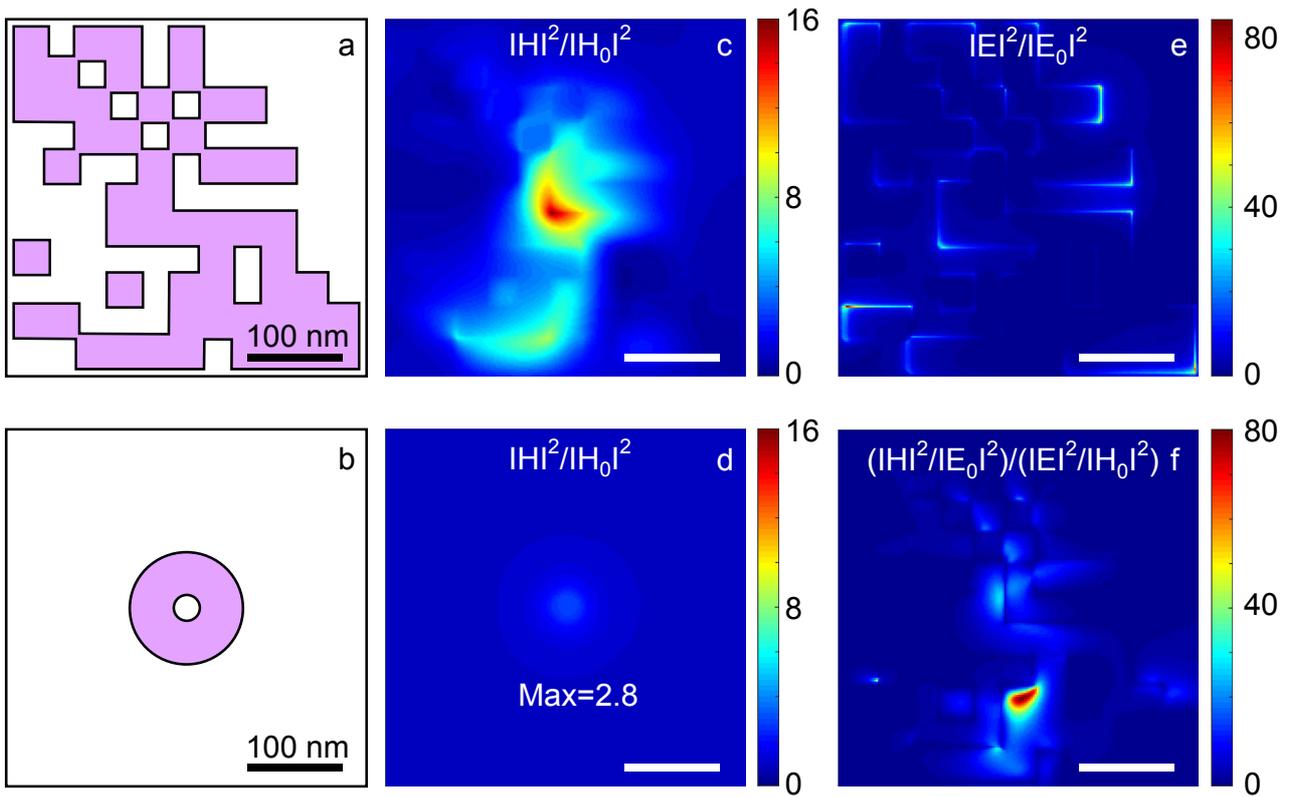

**Figure 3. Optimized structure and comparison with a hollow silicon nanodisk. a) Optimized structure obtained after 248 generations and c) distribution of the magnetic intensity enhancement a) 2nm above the top plane of the antenna in z. b) Schematic of a silicon hollow nanodisk, currently considered as the best dielectric nanoantenna allowing an accessible magnetic hot spot and d) corresponding magnetic intensity increase produced by b), 2nm above the antenna in z as well, and with a maximum of 2.8. e) Electric field intensity enhancement distribution in the same plane as c). f) Magnetic over electric intensity enhancements in the same plane as c) and e).**

Finally, to go further, we investigated the optical behavior of the algorithmically optimized structure with respect to the fine details of its design. Since GA designs are more elaborate than commonly studied optical nanoantennas, their experimental nanofabrication could be challenging. To provide insights towards this issue, we created a new nanostructure, shown in figure 4a, used as an extreme case in which the sharp edges of the design, which are challenging to fabricate, are modified. This rough design follows the general shape of the optimized structure in figure 3a, but instead of using parallelepipeds made of dielectric or air, we perforated a block of silicon with dimensions of 366x366 nm$^2$ in x,y and 92 nm in z, by cylinders made of air with diameters of 38 nm and a height of 92 nm as well. The magnetic intensity distribution produced by this alternative design, potentially easier to fabricate by focused ion beam milling or ebeam lithography, is shown in figure 4b. As we can observe, the spatial distribution of the magnetic intensity is almost not affected by the changes in the structure, and the intensity enhancement is slightly lowered, but is still more than 4 times larger than above a hollow nanodisk (figure 3b and d). This result indicates that once the general GA design is defined, the fine details should only contribute marginally to the field enhancement, increasing the compatibility of the GA designs with experimental nanofabrication.

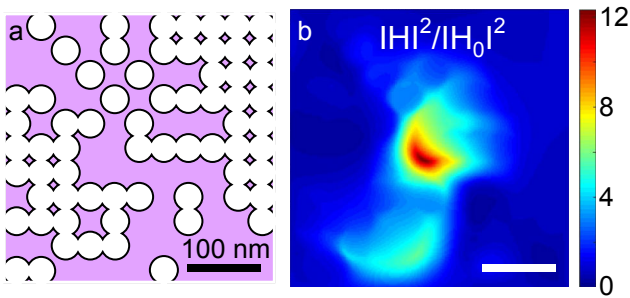

**Figure 4. Low detailed design structure. a) Schematic of an optical nanoantenna where the parallelepipeds of air are replaced by cylinders of 38 nm diameter in the optimized GA design, in order to render the effect of a low detailed structure. b) Magnetic intensity enhancement of a) in a plane 2 nm away in z from the antenna.**

In conclusion, we demonstrated that evolutionary algorithm approaches are a powerful tool for designing dielectric optical nanostructures able to enhance the magnetic intensity of light 5 times more than what state-of-the-art photonic nanoantennas allow nowadays. Furthermore, together with the fine tuning of the optimized design, we demonstrated that although GA designs are more elaborate than most of the optical nanoantennas known so far, the optimized shape is weakly dependent on the fine details (such as sharp edges), and that a rough nanofabrication process would still lead to a much larger enhancement of magnetic light than hollow nanodisks. We believe that this approach will lead to a leap forward in the design of optical nanodevices to study and increase the coupling between magnetic light and matter. In addition, we envision a broad application of the GA approach by setting different selection criteria, such as maximizing the ratio between magnetic over electric intensities, and by using alternative materials or wavelengths. In particular, the use of plasmonic materials should be of interest to enhance the magnetic field at near-infrared telecommunications wavelengths, where erbium ions exhibit strong magnetic dipolar emission.[50]

AUTHOR INFORMATION

**Corresponding Authors**

*E-mail: mathieu.mivelle@insp.upmc.fr

**Author Contributions**

N. B., S. B. And M.M. designed the research. M.M. developed the genetic algorithm and performed the simulations. G.B. developed the FDTD codes. M.M. wrote the manuscript with input from all authors.

The authors declare no competing financial interest.

ACKNOWLEDGMENTS

M.M acknowledges support from the French "Investissements d'Avenir" program (Labex MATISSE), the DIM Nano-K program from "Région Ile de France", the CNRS emergence program and the ANR tremplin under reference ANR-17-ERC3-0006-01.

# Bibliography

1. Landau, L. D.; Lífshíts, E. M.; Pitaevskii, L., *Electrodynamics of continuous media*. Pergamon press Oxford: 1984; Vol. 8.
2. Cowan, R. D., The theory of atomic structure and spectra. Univ of California Press: 1981; Vol. 3.
3. Burresi, M.; Van Oosten, D.; Kampfrath, T.; Schoenmaker, H.; Heideman, R.; Leinse, A.; Kuipers, L., Probing the magnetic field of light at optical frequencies. *Science* **2009,** *326* (5952), 550-553.
4. Giessen, H.; Vogelgesang, R., Glimpsing the weak magnetic field of light. *Science* **2009,** *326* (5952), 529-530.
5. Baranov, D. G.; Savelev, R. S.; Li, S. V.; Krasnok, A. E.; Alù, A., Modifying magnetic dipole spontaneous emission with nanophotonic structures. *Laser & Photonics Reviews* **2017,** *11* (3), 1600268.
6. Valev, V. K.; Govorov, A. O.; Pendry, J., Chirality and Nanophotonics. *Advanced Optical Materials* **2017,** *5* (16).
7. Xi, Z.; Urbach, H., Magnetic Dipole Scattering from Metallic Nanowire for Ultrasensitive Deflection Sensing. *Phys. Rev. Lett.* **2017,** *119* (5), 053902.
8. Manjavacas, A.; Fenollosa, R.; Rodriguez, I.; Jiménez, M. C.; Miranda, M. A.; Meseguer, F., Magnetic light and forbidden photochemistry: the case of singlet oxygen. *Journal of Materials Chemistry C* **2017,** *5* (45), 11824-11831.
9. Noginova, N.; Barnakov, Y.; Li, H.; Noginov, M., Effect of metallic surface on electric dipole and magnetic dipole emission transitions in $Eu^{3+}$ doped polymeric film. *Opt. Express* **2009,** *17* (13), 10767-10772.
10. Karaveli, S.; Zia, R., Strong enhancement of magnetic dipole emission in a multilevel electronic system. *Opt. Lett.* **2010,** *35* (20), 3318-3320.
11. Karaveli, S.; Zia, R., Spectral tuning by selective enhancement of electric and magnetic dipole emission. *Phys. Rev. Lett.* **2011,** *106* (19), 193004.
12. Taminiau, T. H.; Karaveli, S.; van Hulst, N. F.; Zia, R., Quantifying the magnetic nature of light emission. *Nat. Commun.* **2012,** *3*, 979.
13. Karaveli, S.; Wang, S.; Xiao, G.; Zia, R., Time-Resolved Energy-Momentum Spectroscopy of Electric and Magnetic Dipole Transitions in $Cr^{3+}$: MgO. *ACS Nano* **2013,** *7* (8), 7165-7172.
14. Aigouy, L.; Cazé, A.; Gredin, P.; Mortier, M.; Carminati, R., Mapping and quantifying electric and magnetic dipole luminescence at the nanoscale. *Phys. Rev. Lett.* **2014,** *113* (7), 076101.
15. Rabouw, F. T.; Prins, P. T.; Norris, D. J., Europium-Doped NaYF4 Nanocrystals as Probes for the Electric and Magnetic Local Density of Optical States throughout the Visible Spectral Range. *Nano Lett.* **2016,** *16* (11), 7254-7260.
16. Evlyukhin, A. B.; Reinhardt, C.; Seidel, A.; Luk'yanchuk, B. S.; Chichkov, B. N., Optical response features of Si-nanoparticle arrays. *Phys. Rev. B* **2010,** *82* (4), 045404.
17. García-Etxarri, A.; Gómez-Medina, R.; Froufe-Pérez, L. S.; López, C.; Chantada, L.; Scheffold, F.; Aizpurua, J.; Nieto-Vesperinas, M.; Sáenz, J., Strong magnetic response of submicron silicon particles in the infrared. *Opt. Express* **2011,** *19* (6), 4815-4826.
18. Evlyukhin, A. B.; Novikov, S. M.; Zywietz, U.; Eriksen, R. L.; Reinhardt, C.; Bozhevolnyi, S. I.; Chichkov, B. N., Demonstration of magnetic dipole resonances of dielectric nanospheres in the visible region. *Nano Lett.* **2012,** *12* (7), 3749-3755.
19. Rolly, B.; Bebey, B.; Bidault, S.; Stout, B.; Bonod, N., Promoting magnetic dipolar transition in trivalent lanthanide ions with lossless Mie resonances. *Phys. Rev. B* **2012,** *85* (24), 245432.

20. Schmidt, M. K.; Esteban, R.; Sáenz, J.; Suárez-Lacalle, I.; Mackowski, S.; Aizpurua, J., Dielectric antennas-a suitable platform for controlling magnetic dipolar emission. *Opt. Express* **2012,** *20* (13), 13636-13650.
21. Albella, P.; Poyli, M. A.; Schmidt, M. K.; Maier, S. A.; Moreno, F.; Sáenz, J. J.; Aizpurua, J., Low-loss Electric and Magnetic Field-Enhanced Spectroscopy with subwavelength silicon dimers. *J. Phys. Chem. C* **2013,** *117* (26), 13573-13584.
22. Coenen, T.; Van De Groep, J.; Polman, A., Resonant modes of single silicon nanocavities excited by electron irradiation. *ACS Nano* **2013,** *7* (2), 1689-1698.
23. Zambrana-Puyalto, X.; Bonod, N., Purcell factor of spherical Mie resonators. *Phys. Rev. B* **2015,** *91* (19), 195422.
24. Feng, T.; Xu, Y.; Liang, Z.; Zhang, W., All-dielectric hollow nanodisk for tailoring magnetic dipole emission. *Opt. Lett.* **2016,** *41* (21), 5011-5014.
25. Kuznetsov, A. I.; Miroshnichenko, A. E.; Brongersma, M. L.; Kivshar, Y. S.; Luk'yanchuk, B., Optically resonant dielectric nanostructures. *Science* **2016,** *354* (6314), aag2472.
26. van de Haar, M. A.; van de Groep, J.; Brenny, B. J.; Polman, A., Controlling magnetic and electric dipole modes in hollow silicon nanocylinders. *Opt. Express* **2016,** *24* (3), 2047-2064.
27. Feng, T.; Zhang, W.; Liang, Z.; Xu, Y.; Miroshnichenko, A. E., Isotropic Magnetic Purcell Effect. *ACS Photonics* **2017**.
28. Li, J.; Verellen, N.; Van Dorpe, P., Enhancing Magnetic Dipole Emission by a Nano-Doughnut-Shaped Silicon Disk. *ACS Photonics* **2017**.
29. Calandrini, E.; Cerea, A.; De Angelis, F.; Zaccaria, R. P.; Toma, A., Magnetic hot-spot generation at optical frequencies: from plasmonic metamolecules to all-dielectric nanoclusters. *Nanophotonics* **2018**.
30. Zeng, J.; Darvishzadeh-Varcheie, M.; Albooyeh, M.; Rajaei, M.; Kamandi, M.; Veysi, M.; Potma, E. O.; Capolino, F.; Wickramasinghe, H. K., Exclusive Magnetic Excitation Enabled by Structured Light Illumination in a Nanoscale Mie Resonator. *ACS Nano* **2018**.
31. Grosjean, T.; Mivelle, M.; Baida, F.; Burr, G.; Fischer, U., Diabolo nanoantenna for enhancing and confining the magnetic optical field. *Nano Lett.* **2011,** *11* (3), 1009-1013.
32. Hein, S. M.; Giessen, H., Tailoring Magnetic Dipole Emission with Plasmonic Split-Ring Resonators. *Phys. Rev. Lett.* **2013,** *111* (2), 026803.
33. Chigrin, D. N.; Kumar, D.; Cuma, D.; von Plessen, G., Emission Quenching of Magnetic Dipole Transitions near a Metal Nanoparticle. *ACS Photonics* **2015,** *3* (1), 27-34.
34. Mivelle, M.; Grosjean, T.; Burr, G. W.; Fischer, U. C.; Garcia-Parajo, M. F., Strong Modification of Magnetic Dipole Emission through Diabolo Nanoantennas. *ACS Photonics* **2015,** *2* (8), 1071-1076.
35. Hussain, R.; Kruk, S. S.; Bonner, C. E.; Noginov, M. A.; Staude, I.; Kivshar, Y. S.; Noginova, N.; Neshev, D. N., Enhancing Eu 3+ magnetic dipole emission by resonant plasmonic nanostructures. *Opt. Lett.* **2015,** *40* (8), 1659-1662.
36. Ernandes, C.; Lin, H.-J.; Mortier, M.; Gredin, P.; Mivelle, M.; Aigouy, L., Exploring the magnetic and electric side of light through plasmonic nanocavities. *Nano Lett.* **2018,** *18* (8), 5098-5103.
37. Sanz-Paz, M.; Ernandes, C.; Esparza, J. U.; Burr, G. W.; van Hulst, N. F.; Maitre, A. s.; Aigouy, L.; Gacoin, T.; Bonod, N.; Garcia-Parajo, M. F.; Bidault, S.; Mivelle, M., Enhancing Magnetic Light Emission with All-Dielectric Optical Nanoantennas. *Nano Lett.* **2018,** *18* (6), 3481-3487.
38. Wiecha, P. R.; Majorel, C.; Girard, C.; Arbouet, A.; Masenelli, B.; Boisron, O.; Lecestre, A.; Larrieu, G.; Paillard, V.; Cuche, A., Simultaneous mapping of the electric and magnetic photonic local density of states above dielectric nanostructures using rare-earth doped films. *arXiv preprint arXiv:1801.09690* **2018**.
39.     Vaskin, A.; Mashhadi, S.; Steinert, M.; Chong, K. E.; Keene, D.; Nanz, S.; Abass, A.; Rusak, E.; Choi, D.-Y.; Fernandez-Corbaton, I., Manipulation of magnetic dipole emission from Eu3+ with Mie-resonant dielectric metasurfaces. *Nano Lett.* **2019**.
40.     Choi, B.; Iwanaga, M.; Sugimoto, Y.; Sakoda, K.; Miyazaki, H. T., Selective Plasmonic Enhancement of Electric-and Magnetic-Dipole Radiations of Er Ions. *Nano Lett.* **2016,** *16* (8), 5191-5196.
41.     Kuznetsov, A. I.; Miroshnichenko, A. E.; Fu, Y. H.; Zhang, J.; Luk'Yanchuk, B., Magnetic light. *Scientific reports* **2012,** *2*, 492.
42.     Molesky, S.; Lin, Z.; Piggott, A. Y.; Jin, W.; Vuckovic, J.; Rodriguez, A. W., Inverse design in nanophotonics. *Nat. Photonics* **2018,** *12* (11), 659.
43.     Feichtner, T.; Selig, O.; Kiunke, M.; Hecht, B., Evolutionary optimization of optical antennas. *Phys. Rev. Lett.* **2012,** *109* (12), 127701.
44.     Feichtner, T.; Selig, O.; Hecht, B., Plasmonic nanoantenna design and fabrication based on evolutionary optimization. *Opt. Express* **2017,** *25* (10), 10828-10842.
45.     Wiecha, P. R.; Arbouet, A.; Girard, C.; Lecestre, A.; Larrieu, G.; Paillard, V., Evolutionary multi-objective optimization of colour pixels based on dielectric nanoantennas. *Nat. Nanotechnol.* **2017,** *12* (2), 163.
46.     Piggott, A. Y.; Lu, J.; Lagoudakis, K. G.; Petykiewicz, J.; Babinec, T. M.; Vučković, J., Inverse design and demonstration of a compact and broadband on-chip wavelength demultiplexer. *Nat. Photonics* **2015,** *9* (6), 374-377.
47.     Su, L.; Piggott, A. Y.; Sapra, N. V.; Petykiewicz, J.; Vuckovic, J., Inverse design and demonstration of a compact on-chip narrowband three-channel wavelength demultiplexer. *ACS Photonics* **2017,** *5* (2), 301-305.
48.     Yu, Z.; Cui, H.; Sun, X., Genetically optimized on-chip wideband ultracompact reflectors and Fabry–Perot cavities. *Photonics Research* **2017,** *5* (6), B15-B19.
49.     Asano, T.; Noda, S., Optimization of photonic crystal nanocavities based on deep learning. *Opt. Express* **2018,** *26* (25), 32704-32717.
50.     Li, D.; Jiang, M.; Cueff, S.; Dodson, C. M.; Karaveli, S.; Zia, R., Quantifying and controlling the magnetic dipole contribution to 1.5– µ m light emission in erbium-doped yttrium oxide. *Phys. Rev. B* **2014,** *89* (16), 161409.